\documentclass[fleqn,usenatbib]{mnras}

\usepackage{newtxtext,newtxmath}

\usepackage[T1]{fontenc}

\DeclareRobustCommand{\VAN}[3]{#2}
\let\VANthebibliography\thebibliography
\def\thebibliography{\DeclareRobustCommand{\VAN}[3]{##3}\VANthebibliography}

\usepackage{graphicx}	
\usepackage{amsmath}	
\usepackage{hyperref}   
\usepackage{threeparttable}
\usepackage{booktabs}

\title[PNG from DESI Y1 2pcf]{Measuring local primordial non-Gaussianity from the clustering of DESI DR1 LRGs and QSOs}

\author[Z.~Brown et al.]{Z.~Brown$^{1}$\thanks{E-mail: zacherybrown@ksu.edu},
B.~Levi$^{2}$,
H.~Randall$^{2}$,
E.~Chaussidon$^{3}$,
R.~Demina$^{2}$,
A.~G.~Adame$^{39}$,
S.~Avila$^{40}$,
\newauthor
V.~Gonzalez-Perez$^{39,41}$,
J.~Aguilar$^{3}$,
S.~Ahlen$^{4}$,
D.~Bianchi$^{5,6}$,
D.~Brooks$^{7}$,
T.~Claybaugh$^{3}$,
A.~Cuceu$^{3}$,
\newauthor
A.~de la Macorra$^{8}$,
Biprateep~Dey$^{9,10}$,
P.~Doel$^{7}$,
J.~E.~Forero-Romero$^{11,12}$,
E.~Gazta\~naga$^{13,14,15}$,
\newauthor
Satya~Gontcho~A~Gontcho$^{16}$,
G.~Gutierrez$^{17}$,
C.~Hahn$^{18}$,
K.~Honscheid$^{19,20,21}$,
D.~Huterer$^{22,23}$,
M.~Ishak$^{24}$,
\newauthor
R.~Joyce$^{25}$,
D.~Kirkby$^{26}$,
C.~Lamman$^{21}$,
M.~Landriau$^{3}$,
M.~Manera$^{27,28}$,
A.~Meisner$^{25}$,
R.~Miquel$^{29,28}$,
\newauthor
S.~Nadathur$^{14}$,
W.~J.~Percival$^{30,31,32}$,
I.~P\'erez-R\`afols$^{33}$,
A. Ross$^{19}$,
G.~Rossi$^{34}$,
L.~Samushia$^{35,1,36}$,
\newauthor
E.~Sanchez$^{37}$,
M.~Schubnell$^{22,23}$,
J.~Silber$^{3}$,
D.~Sprayberry$^{25}$,
G.~Tarl\'e$^{23}$,
B.~A.~Weaver$^{25}$,
H.~Zou$^{38}$
\\
\\
The authors’ affiliations are listed in Appendix~\ref{app:affiliations}.
}

\date{Accepted XXX. Received YYY; in original form ZZZ}

\pubyear{2025}

\begin{document}
\label{firstpage}
\pagerange{\pageref{firstpage}--\pageref{lastpage}}
\maketitle

\begin{abstract}

We report the first measurement of primordial non-Gaussianity (PNG), parameterized by $f_{\mathrm{NL}}$, in the configuration space two-point correlation function (2pcf). We employ simulation based modeling and a novel approach for the mitigation of imaging systematics. We apply this method to samples of luminous red galaxies (LRG) and quasars (QSO) observed by the Dark Energy Spectroscopic Instrument (DESI) during the first year of its observations (DR1). The observed 68\% CL interval on $f_{\mathrm{NL}}$ is $-3^{+22}_{-21}$ using LRGs, and $ 0^{+17}_{-16}$ using QSOs. The joint measurement yields $f_{\mathrm{NL}} = -3^{+12}_{-12}$ at $\ [68\%]$ CL. Our pipeline imposes a Gaussian prior on the value of $p$ (which defines the PNG bias via the Universality relation), with $p_{\rm LRG} = 1.0 \pm 0.1$ and $p_{\rm QSO} = 1.6\pm 0.1$. The observed constraining power of DESI tracers significantly exceeds that of previous large-scale structure (LSS) surveys, and encouragingly, approaches the sensitivity of CMB probes of PNG.

\end{abstract}




\section{Introduction}
\label{sec:intro}

Models of inflation describe an epoch during which the early Universe underwent rapid, exponential expansion. The nature of this process remains under active investigation in contemporary cosmology. Absent the ability to probe the fields driving inflation directly, the most promising tests infer their properties from the clustering of matter at later epochs. In surveys of large-scale structure (LSS), primordial non-Gaussianity (PNG) is constrained by its effects on galaxy clustering statistics.

In the presence of PNG, the primordial potential $\Phi$ deviates quadratically from that of a Gaussian random field, $\phi$, using a single free parameter $f_{\mathrm{NL}}$:
\begin{equation}
\label{eq:primordial_potential}
\Phi = \phi + f_{\mathrm{NL}} \left( \phi^2 -\langle \phi \rangle^2 \right) \ ,
\end{equation}
where $\phi$ is the potential of a Gaussian random field \citep{gangui1993three,komatsu2001acoustic,komatsu2002cosmic,mueller2019optimizing}. The magnitude of the quadratic term, $f_{\mathrm{NL}}$, is then used to distinguish between inflationary models. Small or zero values ($ |f_{\mathrm{NL}}| \lesssim O(0.1)$) indicate the presence of a single scalar field during inflation, the so-called ``slow-roll'' model \citep{maldacena2003non, bartolo2004non}. Conversely, detection of large non-zero $f_{\mathrm{NL}}$ would indicate a more complex, multi-field model \citep{creminelli2011not}. For decades, temperature fluctuations of the cosmic microwave background (CMB) have been the most sensitive probe of PNG, with measurements from Planck yielding $f_{\mathrm{NL}} = -0.9 \pm 5.1$ \citep{collaboration2020planck}. 

We are now, however, entering into an era where the volume and density of Large Scale Structure (LSS)  surveys are expected to deliver a competitive constraining power on PNG \citep{mueller2021clustering,chaussidon2025constrainingprimordialnongaussianitydesi}, if the systematic uncertainties due to imaging quality can be kept under control. Here, we present a measurement of $f_{\mathrm{NL}}$, performed using the configuration-space two-point correlation function (2pcf) of two classes of tracer from the first data release (DR1) of the Dark Energy Spectroscopic Survey (DESI): the luminous red galaxies (LRG) and quasars (QSO)  \citep{desicollaboration2025datarelease1dark}. DESI is a fiber-fed spectroscopic surveyor mounted on the Mayall 4-meter telescope at Kitt Peak National Observatory \citep{2022AJ....164..207A,2023arXiv230606310M,2023AJ....165....9S,2016arXiv161100037D,2016arXiv161100036D}. Tasked with uncovering the nature of Dark Energy \citep{2013arXiv1308.0847L}, DESI has been observing galaxy spectra across a third of the sky at a pace of almost 5000 spectra per exposure. This large volume survey is also ideal for studies of PNG. The DESI program observes a large collection of target spectra, from local stars to high redshift quasars \citep{2020RNAAS...4..188A,2020RNAAS...4..187R,2020RNAAS...4..181Z,2020RNAAS...4..180R,2020RNAAS...4..179Y,2023ApJ...943...68L,2023AJ....165..124A,2022arXiv220808514C,2022arXiv220808512H,2023AJ....165...58Z,2023AJ....165..126R,2023ApJ...944..107C}. In this work, we restrict our analysis to the LRG and QSO catalogs, which for DR1 cover approximately half of the planned main survey.

Studies of PNG from LSS are notoriously challenging due to the potential contamination by imaging systematics \citep{huterer2013calibration}. Beyond the simple incompleteness of the DR1 catalogs, effects like dust extinction, stellar density, seeing, and survey depth all induce variations in angular survey density at large scales \citep{rezaie2024local}. In the galaxy 2pcf, this can cause an excess of large scale clustering that mimics the effect of PNG \citep{thomas2011excess}. Many strategies have been developed to account for imaging systematics in measurements of PNG, for example in \cite{chaussidon2025constrainingprimordialnongaussianitydesi}, and  \cite{brown2024constrainingprimordialnongaussianitylarge} used in this work.

The paper is organized as follows. In Sec.~\ref{sec:algorithm_details} we describe the algorithm used to estimate the galaxy 2pcf, followed by an overview of the statistical model in Sec.~\ref{sec:theory_model}. The measurements of $f_{\mathrm{NL}}$ are presented in Sec.~\ref{sec:desi_measurement}, after which we discuss their implications and interpretations in Sec.~\ref{sec:discussion}, and conclude in Sec.~\ref{sec:conclusions}.


\section{Evaluation of the 2pcf} 
\label{sec:algorithm_details}

The galaxy 2pcf is estimated using the {\tt ConKer} algorithm \citep{brown2022conker}. Following the work of \cite{brown2021algorithm}, this method convolves the matter density field with spherical shell kernels to achieve $\mathcal{O}(N_g)$ 2pcf computations, where $N_g$ denotes the number of tracers. Below, we briefly describe the procedure employed by the {\tt ConKer} algorithm.

 We evaluate the comoving radial distance, $r$ from the observed redshift, $z$ of each tracer using Hubble integral 
\begin{equation}
\label{eq:hubbleint}
r(z) = \frac{c}{H_0} { \displaystyle\int_{0}^{z} } \frac{dz'}{ \sqrt{\Omega_M(z'+1)^3+\Omega_k(z'+1)^2+\Omega_\Lambda} }\ , 
\end{equation} 
where $\Omega_M$, $\Omega_k$, and $\Omega_\Lambda$ are the relative present day matter, curvature, and cosmological constant densities respectively. $H_0=h \times 100$ km s$^{-1}$ Mpc$^{-1}$ is the present day Hubble's constant, and $c$ is the speed of light. In this analysis we choose a flat $\Lambda$CDM cosmology with $\Omega_{M} = 0.315$. Comoving distances are computed in units of $h^{-1}$Mpc. 
We begin with a catalog of tracers $D$, and corresponding randoms $R$. We cast each tracer onto a three-dimensional cartesian grid $D(X,Y,Z)$ and $R(X,Y,Z)$. When defining our grids $D(X,Y,Z)$ and $R(X,Y,Z)$, tracers are assigned to each cell with a weight, given by 
\begin{equation}
\label{eq:weights}
w_{\mathrm{tot}} = w_{\mathrm{sys}} \times w_{\mathrm{comp}} \times w_{\mathrm{zfail}} \times w_{\mathrm{FKP}}\ . 
\end{equation} 
A correction for imaging systematics is provided by $w_{\mathrm{sys}}$, the incompleteness of the DR1 survey catalogs is mitigated by $w_{\mathrm{comp}}$, and catastrophic failures of the redshift pipeline are accounted for by $w_{\mathrm{zfail}}$. The only redshift dependent weight is $w_{\mathrm{FKP}}$, which prioritize tracers at underpopulated redshift values, based on their distribution in $z$, $n(z)$  \citep{feldman1993power}. Thus, the weighted number of objects in a given catalog is given by the sum of these weights,
\begin{equation}
\label{eq:ntot}
N_{\mathrm{tot}} = \sum_{i}^{N_g} w_{\mathrm{tot},i} \ . 
\end{equation}
The density contrast field, $C(X,Y,Z)$ is the normalized difference between the data and random fields,
\begin{equation}
\label{eq:dens_contrast}
C(X,Y,Z) = D(X,Y,Z) - \frac{N_{\mathrm{tot},D}}{N_{\mathrm{tot},R}}R(X,Y,Z) \ , 
\end{equation}
where $N_{\mathrm{tot},D}$ and $N_{\mathrm{tot},R}$ are the weighted number of tracers in the data and randoms catalogs. For each scale $s$ in the 2pcf, a kernel grid $K_s(X,Y,Z)$ is defined such that a cell within a distance $s\pm\Delta s / 2$ from the center is assigned a value of unity. $\Delta s$ is the width of each 2pcf bin. All other cells are assigned a value of zero. Should a cell partially overlap a spherical shell corresponding to some $s$, that cell is assigned a value between 0 and 1, representing the fraction of its volume that falls within $s\pm\Delta s / 2$ from the center. From these grids, we define the monopole of the 2pcf, to be referred to as $\xi(s)$, 
\begin{equation}
\label{eq:2pcf_monopole}
\xi(s) = \frac{\sum_{\mathrm{grid}} C \times (K_s \ast\ast\ast C)}{\sum_{\mathrm{grid}} R \times (K_s \ast\ast\ast R)} \ , 
\end{equation}
where the operation $\ast\ast\ast$ is a discreed three-dimensional convolution, and the sum is performed over all cells in the grid.

In this study, we measure all 2pcfs using bins of width $\Delta s = 10$ $h^{-1}$Mpc from $s = 50\  h^{-1}\mathrm{Mpc}\ \mbox{---}\ 380$ $h^{-1}$Mpc. For both the LRG and QSO samples, the survey is split into Northern Galactic Cap (NGC) and Southern Galactic Cap (SGC) tracers. To present the complete DESI DR1 2pcf for each tracer, we combine the NGC and SGC by adding the numerator and denominator of Eqn.~\ref{eq:2pcf_monopole} for each region.


\section{Description of the model}
\label{sec:theory_model}

We employ a statistical model first described in \cite{brown2024constrainingprimordialnongaussianitylarge}. It is entirely simulation based. We start with a fiducial model without PNG ($f_{\mathrm{NL}} = 0$) based on an ensemble of AbacusSummit high resolution N-body simulations \citep{maksimova2021abacussummit,garrison2021abacus}, distributed about halos using a halo occupation distribution (HOD) model defined in \cite{bose2022constructing,yuan2023full}. The fiducial 2pcf, $\xi^{\mathrm{fid}}(s)$, is an average of the 25 realizations with cutsky survey geometry, matching the number density and volume of the DESI DR1 LRG and QSO catalogs.

The effect of PNG on the 2pcf monopole is modeled using a suite approximated N-body, or ``FastPM'' simulations of dark matter halos \citep{feng2016fastpm}.  In this study we use 100 realizations of $(3\ h^{-1}\mathrm{Gpc})^3$ boxes with fixed, paired, and matched initial conditions at both $f_{\mathrm{NL}} = 0$ and $f_{\mathrm{NL}} = 100$. Both cases are modeled from cubic boxes evolved to redshift $z$ = 1. The dependence of 2pcf, $\xi(s,f_{\mathrm{NL}})$  on $f_{\mathrm{NL}}$ is linearly interpolated using coefficients $A(s)$ between these two sets:
\begin{equation}
\label{eq:A_fnl}
\left[ \xi(s,f_{\mathrm{NL}}) - \xi(s,0) \right]= A(s) f_{\mathrm{NL}} \ .
\end{equation}

From these, the model 2pcf may be decomposed into two terms as follows,
\begin{equation}
\label{eq:png_2pcf_theory}
\xi(s, f_{\mathrm{NL}}) =  \left( \frac{\tilde{b}_{1}}{\tilde{b}_{1}^{\mathrm{fid}}} \right)^2 \xi^{\mathrm{fid}}(s) + \left( \frac{\tilde{b}_{1}}{b_{1h}}\right) ^2 A(s) r_{\phi} f_{\mathrm{NL}} \ . 
\end{equation}
Here, $\tilde{b}_{1}$ refers to the linear galaxy bias modified by the Kaiser term to account for redshift-space distortions \citep{kaiser1987clustering}. $\tilde{b}_{1}^{\mathrm{fid}}$ is the same linear bias for the fiducial (no PNG) simulation. In the absence of PNG, the first term in Eqn.~\ref{eq:png_2pcf_theory} simply scales $\xi(s, f_{\mathrm{NL}})$ to $\xi^{\mathrm{fid}}(s)$ according to the ratio of the linear biases squared.  The effect of PNG is encoded in the second term. The bias ratio describes the scaling of halos described by halo bias, $b_{1h}$  to the model tracer with the corresponding linear bias $\tilde{b}_{1}$.  
The presence of PNG modifies the density fluctuations according to the product of $f_{\mathrm{NL}}$ and $b_{\phi}$, the PNG bias, which encodes the response of a tracer to the primordial potential, and varies for different tracer types. The relation between $b_1$ and $b_{\phi}$  is derived from the Universality relation \citep{barreira2020impact} 
\begin{equation}
b_{\phi }=2\delta_c (b_{1}-p) \ ,
\label{eq:bphi}
\end{equation}
where $\delta_c=1.69$ is the critical density of spherical collapse, and $p$ is the tracer-dependent parameter describing the formation of said tracer in primordial overdensities. 
In Eqn.~\ref{eq:png_2pcf_theory}, the value of $r_{\phi}$ accounts of the difference in $b_{\phi}$ for the simulated halos and the tracer: 
\begin{equation}
\label{eq:Rphi_def}
r_{\phi} = \frac{b_{1 h}}{\tilde{b}_{1}} \frac{b_{\phi }}{b_{\phi h}}   = \frac{b_{1 h}}{\tilde{b}_{1}} \frac{b_{1} - p}{b_{1h} - p_h}\ , 
\end{equation}
where $p$ refers to the model tracer, and $p_h$ to halos. 

The model in Eqn.~\ref{eq:png_2pcf_theory} is an idealized case without survey systematics. For this work, however, it is critical that we account for spurious effects arising from imaging systematics. If the values of $w_{\mathrm{sys}}$ are over-estimated or under-estimated, it will result in an excess or deficit of clustering in the 2pcf. We model the deviations in the 2pcf due to variation in the values of $w_{\mathrm{sys}}$ using a simulation based approach. We apply the same imaging weights to the AbacusSummit fiducial simulations as are assigned to the true DESI tracers using the software package {\tt regressis}, according to \cite{chaussidon2022angular}. DESI photometry is derived from three main survey regions \citep{silva2016mayall,flaugher2015dark}. The NGC is split into the Mayall z-band Legacy Survey (MzLS) and the Dark Energy Camera Legacy Survey (DECaLS), with MzLS covering the high declination angles and DECaLS covering nearer to the equator. The SGC is also covered by DECaLS imaging. Since these regions have different image quality we include three independent parameters, $K_{\mathrm{sys}}^{R}$, where $R$ = MzLs, (N)DEC, (S)DEC, which correspond to the amount by which the weights are varied in a given region. 
The value of $K_{\mathrm{sys}}^{R}$ corresponds to a percent by which the weights are over or under-applied, where the resultant weight for an object, $w_{\mathrm{sys},i}^{*}$, is given by
\begin{equation}
  w_{\mathrm{sys},i}^{*} = 1 + \left( 1+ \frac{K^{R}_{\mathrm{sys}}}{100} \right) (w_{\mathrm{sys},i} - 1)  \ .
\label{eq:ksys}
\end{equation}
In this parameterization, $K^{R}_{\mathrm{sys}} = -25$, for example, is interpreted as a $25\%$ under-application of the weights in region $R$. 

Using our simulated DR1 catalogs, we evaluate the effect of each imaging weight independently on the 2pcf of the entire sample. For each region, the process is repeated for five values of $K^{R}_{\mathrm{sys}} = -50, -25, 0, 25, 50$ on the AbacusSummit DR1 simulations. The dependence of the 2pcf versus $K^{R}_{\mathrm{sys}}$ is fit to a quadratic function in each bin. Since $K^{R}_{\mathrm{sys}} = 0$ represents the fiducial case with no deviation, the quadratic fit is composed of two parameters, $\mathcal{C}^{R}$ and $\mathcal{D}^{R}$. Eqn.~\ref{eq:png_2pcf_theory} then gets an additional term describing the variation due to imaging systematics in each observational region:
\begin{equation}
    \sum_R \left( \frac{\tilde{b}_{1}}{\tilde{b}_{1}^{\mathrm{fid}}} \right)^2 \left[ \mathcal{C}^{R} (K^{R}_{\mathrm{sys}})^2 + \mathcal{D}^{R} (K^{R}_{\mathrm{sys}}) \right]  \ .
\label{eq:ksys_term}
\end{equation}

We then construct a likelihood function corresponding to this model. The observable $O$ is the 2pcf monopole of either the LRG or QSO samples. The expected (or model) 2pcf is given by $\tilde{O}$. The model depends on two parameters of interest (POIs): $f_{\mathrm{NL}}$ and the linear bias $b_{1}$, and on seven nuisance parameters $\theta$:
\begin{equation}
\label{eq:params_fid}
\theta = [b_{1h}, b_{1}^{\mathrm{fid}}, p_h, p, K^{\mathrm{MzLS}}_{\mathrm{sys}}, K^{\mathrm{(N)DEC}}_{\mathrm{sys}}, K^{\mathrm{(S)DEC}}_{\mathrm{sys}}]  \ .
\end{equation}
In this analysis all nuisance parameters are constrained using Gaussian priors where the central values are given by a vector $\Theta$ and Gaussian widths $\sigma$. The test statistic is 
\begin{equation}
\chi^2 = \frac{1}{2} \sum_{ij}\left[O_i-\tilde{O}_i\right]^TC^{-1}_{ij}\left[O_j-\tilde{O}_j\right] + \frac{1}{2}\sum_f \frac{(\theta_f - \Theta_f)^2}{(\sigma_f)^2}\ , 
\label{eq:chi2}
\end{equation}
where $C_{ij}$ denotes the covariance matrix. The first summation over indices $i$ and $j$ is performed over bins in $s$ and the second over nuisance parameters $\theta_f$. $\chi^2$ is sampled to find the optimal values of the POIs for both the LRG and QSO samples. Our covariance matrices are derived from suites of 1000 mocks based on the effective Zel'dovich (EZ) approximation \citep{chuang2015ezmocks}. The EZ mocks used for covariance also cover the same geometry and match density with the true DESI samples.

\section{Application to DESI galaxies} 
\label{sec:desi_measurement}

Here, we present our measurements of $f_{\mathrm{NL}}$ first from the DR1 LRG sample, then from the DR1 QSO sample. We also combine the resultant posterior distributions to provide a joint estimate of $f_{\mathrm{NL}}$.

\subsection{LRG Sample}
\label{subsec:lrg_details}

\begin{figure}
\includegraphics[width=\columnwidth]{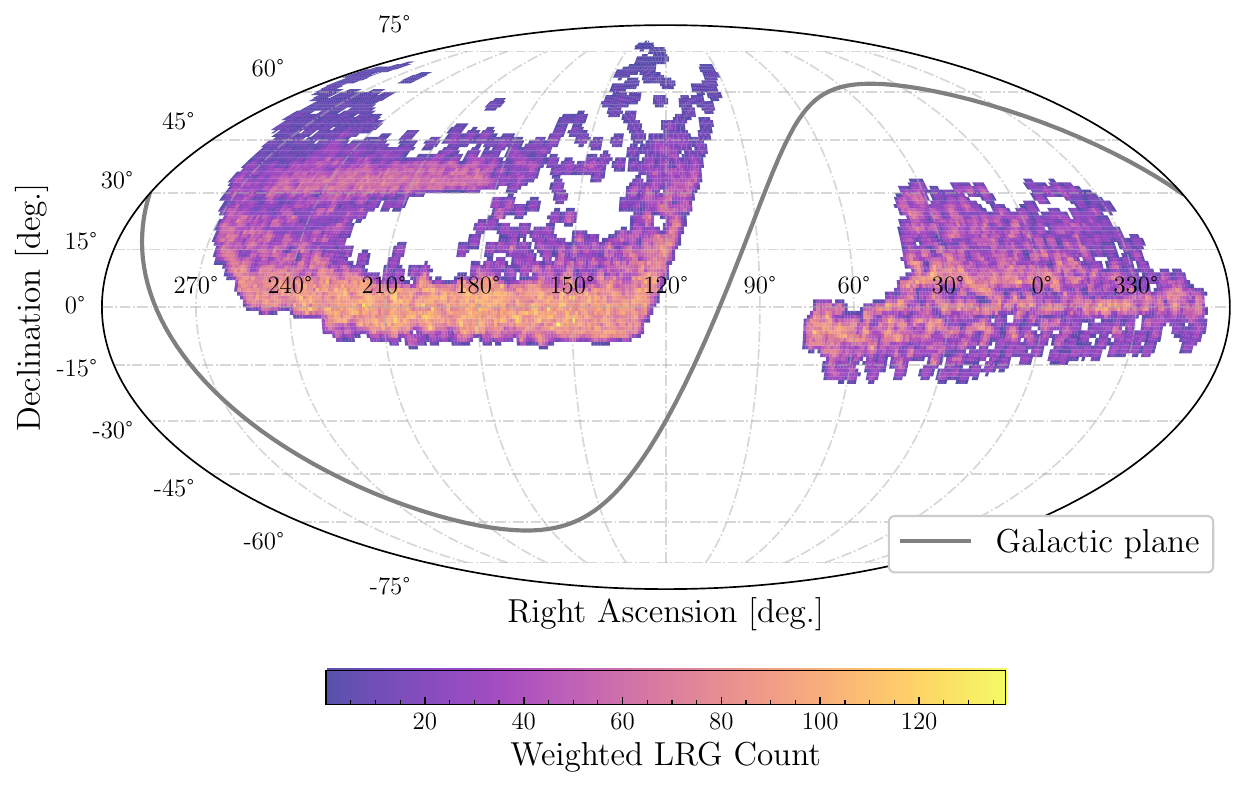}
\caption{Weighted angular density distributions of the DESI Y1 LRG samples (colorscale), showing both the NGC and SGC separated by the galactic plane (grey band). LRG counts are weighted by $w_{\mathrm{tot}}$.}
\label{fig:sky_hist_lrg}
\end{figure}

\begin{figure}
\includegraphics[width=\columnwidth]{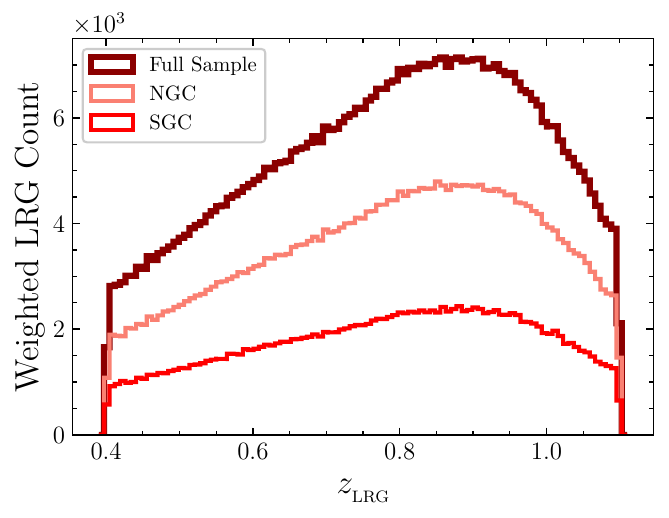}
\caption{Weighted redshift distributions of the DESI Y1 LRG samples, showing the NGC (pink), SGC (red), and combined (dark red). LRG counts are weighted by $w_{\mathrm{tot}}$.}
\label{fig:z_hist_lrg}
\end{figure}

The LRG catalogs employed in this study consist of 1 511 049 NGC tracers and 686 696 SGC tracers spanning a redshift range from 0.4 $<$ $z$ $<$ 1.1. The weighted effective redshift for the complete sample is $z_{\mathrm{eff}} = 0.781$. The angular distribution of LRGs is shown in Fig.~\ref{fig:sky_hist_lrg}, and the redshift distribution is shown in Fig.~\ref{fig:z_hist_lrg}. In both histograms, we present the weighted number of tracers. 

\subsection{LRG Measurement}
\label{subsec:lrg_fits}

\begin{figure}
\includegraphics[width=\columnwidth]{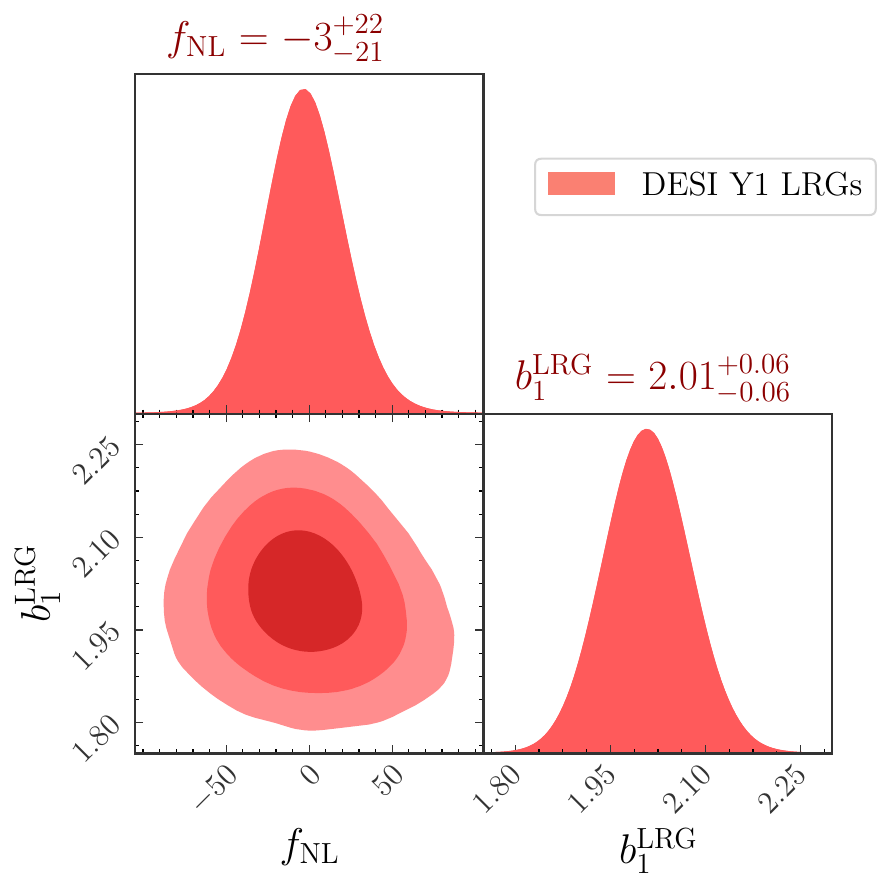}
\caption{Marginalized distributions of the POIs for DESI Y1 LRGs.
The dark to light regions represent $1\sigma$, $2\sigma$, and $3\sigma$ contours. The most probable values and $1\sigma$ CL are labeled for each parameter above the respective panel.}
\label{fig:post_lrg}
\end{figure}

\begin{figure}
\includegraphics[width=\columnwidth]{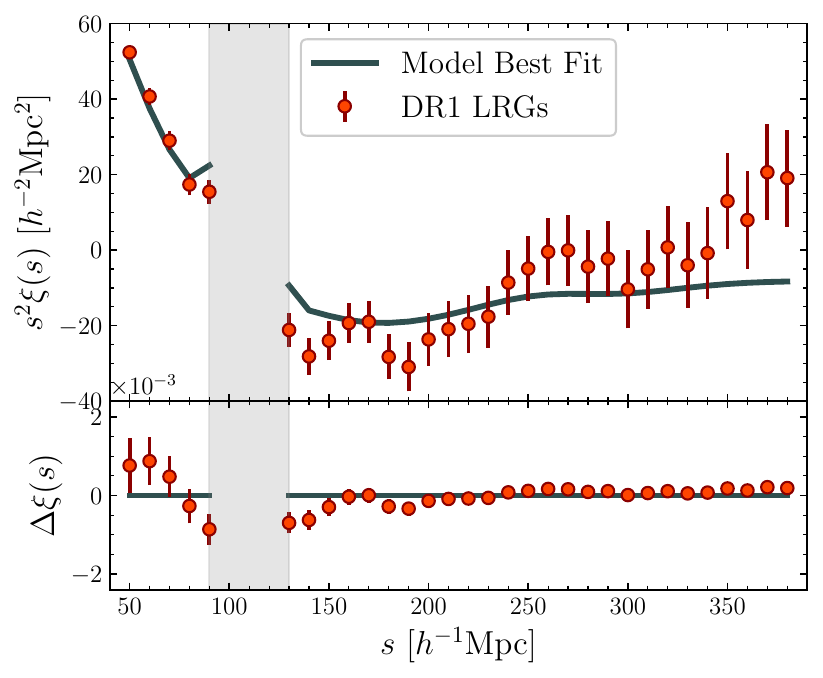}
\caption{The upper panel shows the 2pcf of DR1 DESI LRGs (red markers), with error bars derived from 1000 EZ mocks and the model 2pcf corresponding to the best fitting parameters (grey solid line). The bottom panel shows the residual between the data and best fitting model. The shaded band represents the BAO scales which have been masked out in this analysis.}
\label{fig:best_fit_lrg}
\end{figure}

\begin{table}
\begin{center}
\begin{threeparttable}
\begin{tabular}{lll}
\toprule
Parameter & LRG & QSO \\
\midrule
$f_{\mathrm{NL}}$ & $\mathcal{U}[-250,250]$ & $\mathcal{U}[-250,250]$ \\
$b_1$ & $\mathcal{U}[0.5,4]$ & $\mathcal{U}[0.5,4]$ \\
$b_{1h}$ & $\mathcal{N}(2.67,0.04)$ & $\mathcal{N}(2.67,0.04)$ \\
$b_{1}^{\mathrm{fid}}$ & $\mathcal{N}(1.94,0.055)$ & $\mathcal{N}(2.66,0.055)$ \\
$p_h$ & $\mathcal{N}(1.0,0.1)$ & $\mathcal{N}(1.0,0.1)$ \\
$p$ & $\mathcal{N}(1.0,0.1)$ & $\mathcal{N}(1.6,0.1)$ \\
$K^{\mathrm{MzLS}}_{\mathrm{sys}}$ & $\mathcal{N}(0.0,10.0)$ & $\mathcal{N}(0.0,10.0)$ \\
$K^{\mathrm{(N)DEC}}_{\mathrm{sys}}$ & $\mathcal{N}(0.0,10.0)$ & $\mathcal{N}(0.0,10.0)$ \\
$K^{\mathrm{(S)DEC}}_{\mathrm{sys}}$ & $\mathcal{N}(0.0,10.0)$ & $\mathcal{N}(0.0,10.0)$ \\
\bottomrule
\end{tabular}
\caption{Priors applied to the parameters of the fit for LRGs (left) and QSOs (right). $\mathcal{U}$ indicates a flat prior while $\mathcal{N}$ denotes a Gaussian prior.}
\vspace{-0.2cm}
\label{tab:priors}
\end{threeparttable}
\end{center}
\end{table}

\begin{table}
\begin{center}
\begin{threeparttable}
\renewcommand{\arraystretch}{1.5}
\begin{tabular}{lll}
\toprule
Parameter & LRG & QSO \\
\midrule
$f_{\mathrm{NL}}$ & $-3^{+22}_{-21}$ & $0^{+17}_{-16}$ \\
$b_1$ & $2.01^{+0.06}_{-0.06}$ & $2.57^{+0.10}_{-0.10}$ \\
$b_{1h}$ & $2.68^{+0.03}_{-0.03}$ & $2.68^{+0.03}_{-0.03}$ \\
$b_{1}^{\mathrm{fid}}$ & $1.93^{+0.06}_{-0.06}$ & $2.66^{+0.04}_{-0.04}$ \\
$p_h$ & $1.0^{+0.1}_{-0.1}$ & $1.0^{+0.1}_{-0.1}$ \\
$p$ & $1.0^{+0.1}_{-0.1}$ & $1.6^{+0.1}_{-0.1}$ \\
$K^{\mathrm{MzLS}}_{\mathrm{sys}}$ & $-0.3^{+9.9}_{-9.9}$ & $1.2^{+9.5}_{-9.6}$ \\
$K^{\mathrm{(N)DEC}}_{\mathrm{sys}}$ & $0.1^{+10.0}_{-10.0}$ & $0.3^{+9.9}_{-9.9}$ \\
$K^{\mathrm{(S)DEC}}_{\mathrm{sys}}$ & $0.0^{+9.9}_{-9.9}$ & $-1.0^{+9.8}_{-9.8}$ \\
\bottomrule
\end{tabular}
\caption{The best fitting values for each parameter of our fits, expressed as 68\% confidence limits resulting from our measurements.}
\vspace{-0.2cm}
\label{tab:posteriors}
\end{threeparttable}
\end{center}
\end{table}

While performing the PNG measurement on the LRG sample, we assign the following Gaussian priors to the first four nuisance parameters: $b_{1h} = 2.67 \pm 0.04$, $p_h = 1.0 \pm 0.1$, $b_{1}^{\mathrm{fid}} = 1.94 \pm 0.055$, $p = 1.0 \pm 0.1$. For the imaging nuisance parameters, we assign each a value of $ K^{R}_{\mathrm{sys}} = 0.0 \pm 10.0$, which allows a 10$\%$ variation in the values of the imaging weights in each region, centered on their fiducial values. The POIs $f_{\mathrm{NL}}$ and $b_1$ are unconstrained. A summary of the priors applied to the LRG fit is given in Tab.~\ref{tab:priors}. The test statistic is sampled by a Markov Chain Monte-Carlo (MCMC) algorithm with 75 walkers each executing 20K steps. In our analysis, we mask and exclude the BAO scales while performing the fit. Specifically, we remove 2pcf measurement that fall between $s$ = 100 $h^{-1}$ Mpc and 120 $h^{-1}$ Mpc. Since the BAO peak is subject to non-linear clustering effects which are not described by our statistical model, the inclusion of those scales may bias our result. A detailed discussion of the BAO masking procedure and motivations may be found in App.~\ref{app:bao_mask}.

We show the marginalized distributions of $f_{\mathrm{NL}}$ and $b_1$ in Fig.~\ref{fig:post_lrg}. Here, we find that $f_{\mathrm{NL}}$ is consistent with zero, with 1$\sigma$ uncertainties of approximately 22. In Fig.~\ref{fig:best_fit_lrg}, we plot the 2pcf for the combined NGC and SGC sample, and show the model corresponding to the best fitting set of parameters. The best fitting values for all parameters in the fit are presented in Tab.~\ref{tab:posteriors}. The nuisance parameters are nearly entirely constrained by the priors imposed upon them.

\subsection{QSO Sample}
\label{subsec:qsodetails}

\begin{figure}
\includegraphics[width=\columnwidth]{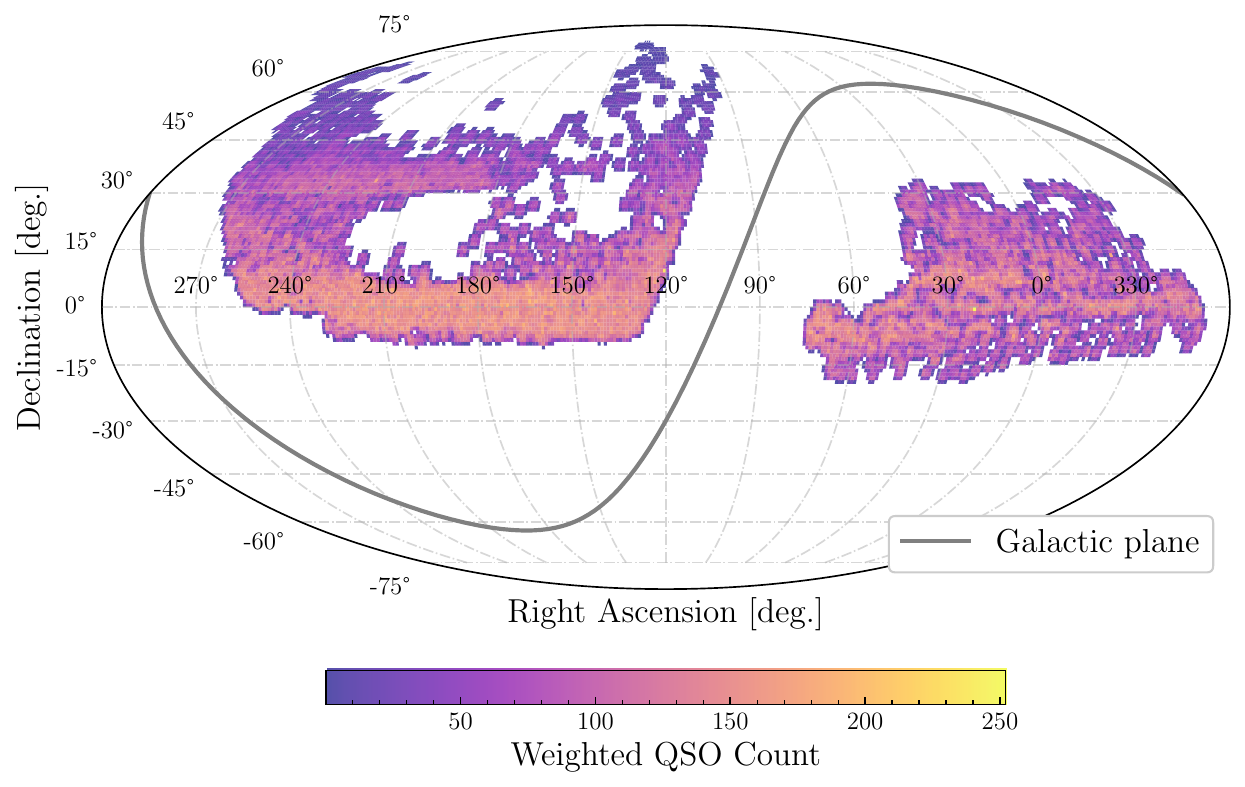}
\caption{Weighted angular density distributions of the DESI Y1 QSO samples (colorscale), showing both the NGC and SGC separated by the galactic plane (grey band). QSO counts are weighted by $w_{\mathrm{tot}}$.}
\label{fig:sky_hist_qso}
\end{figure}

\begin{figure}
\includegraphics[width=\columnwidth]{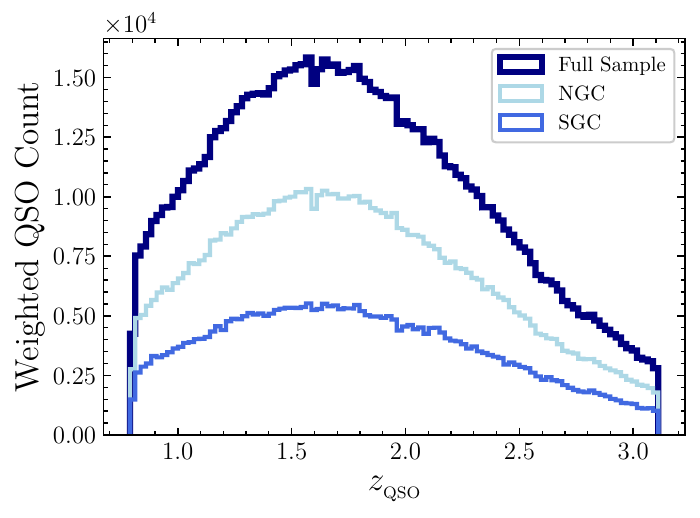}
\caption{Weighted redshift distributions of the DESI Y1 QSO samples, showing the NGC (light blue), SGC (blue), and combined (dark blue). QSO counts are weighted by $w_{\mathrm{tot}}$.}
\label{fig:z_hist_qso}
\end{figure}

We use QSO catalogs of 793,229 NGC tracers and 430,172 SGC tracers covering a redshift range from 0.8 $<$ $z$ $<$ 3.1. The weighted effective redshift for the complete sample is $z_{\mathrm{eff}} = 1.832$. In Fig.~\ref{fig:sky_hist_qso}, we show the weighted angular distribution of QSOs followed by the weighted redshift distribution of this sample in Fig.~\ref{fig:z_hist_qso}.

\subsection{QSO Measurement}
\label{subsec:qso_fits}

\begin{figure}
\includegraphics[width=\columnwidth]{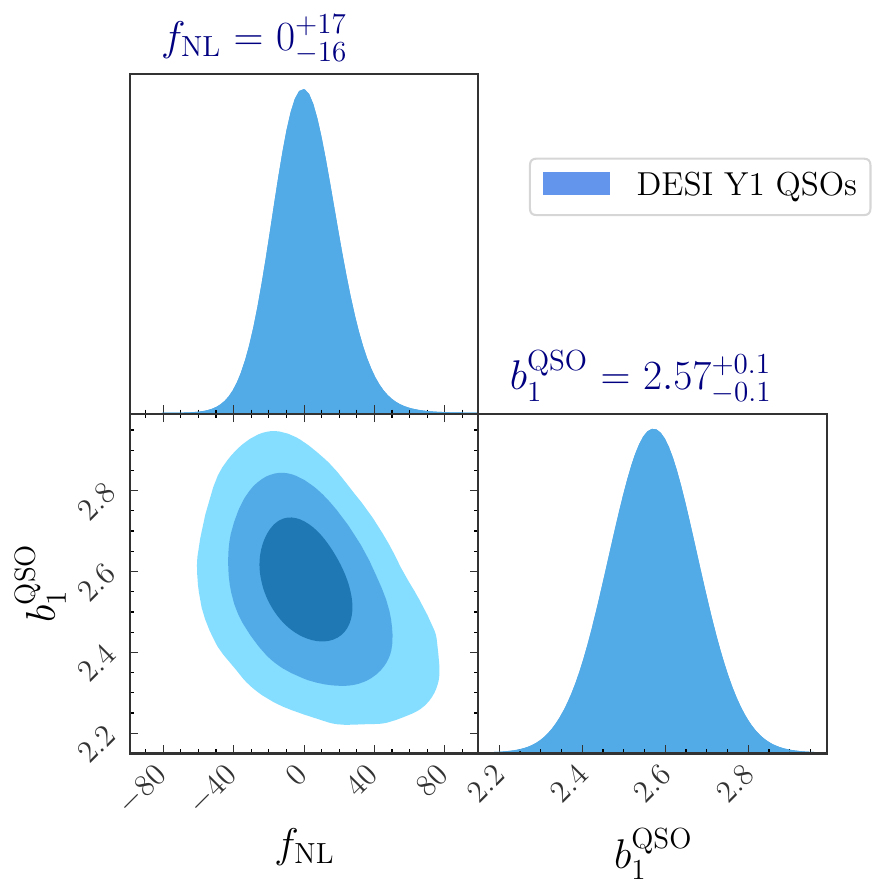}
\caption{The same as Fig.~\ref{fig:post_lrg} for the DESI DR1 QSO sample.}
\label{fig:post_qso}
\end{figure}

\begin{figure}
\includegraphics[width=\columnwidth]{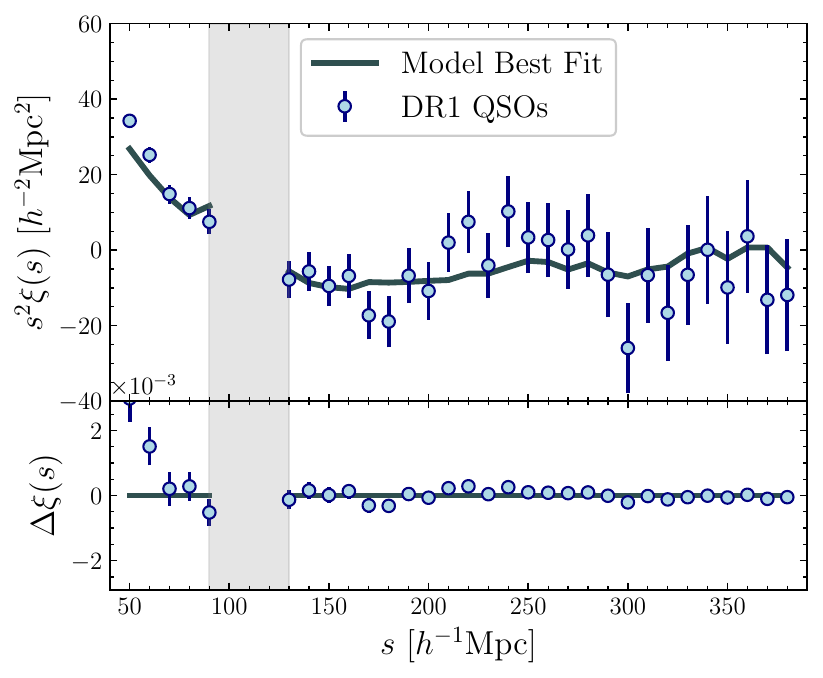}
\caption{The same as Fig.~\ref{fig:best_fit_lrg} for the DESI DR1 QSO sample.}
\label{fig:best_fit_qso}
\end{figure}

The following Gaussian priors are assigned to the first four nuisance parameters: $b_{1h} = 2.67 \pm 0.04$, $p_h = 1.0 \pm 0.1$, $b_{1}^{\mathrm{fid}} = 2.66 \pm 0.055$, $p = 1.6 \pm 0.1$. Of particular note here is the change in the value of $p$ with respect to the LRGs. Studies have shown that a larger value is appropriate for these tracers \citep{adame2024png}. As with the LRGs, the imaging priors are all $ K^{R}_{\mathrm{sys}} = 0.0 \pm 10.0$. The details of the priors applied to the QSO fits may be found in Tab.~\ref{tab:priors}. The test statistic is sampled by an MCMC algorithm with the same parameters as were used for the LRGs. The same BAO scales are also excluded in the QSO fit.

We show the marginalized distributions of $f_{\mathrm{NL}}$, $b_1$ in Fig.~\ref{fig:post_qso}. Here, we find that $f_{\mathrm{NL}}$ is again consistent with zero, this time with 1$\sigma$ uncertainties of approximately 16. In Fig.~\ref{fig:best_fit_qso}, we plot the 2pcf for the combined NGC and SGC sample, and show the model corresponding to the best fitting set of parameters. We summarize the best fitting values for the fit parameters in the QSO measurement in Tab.~\ref{tab:posteriors}.

\subsection{Joint Fits}
\label{subsec:joint_fits}

\begin{figure}
\includegraphics[width=\columnwidth]{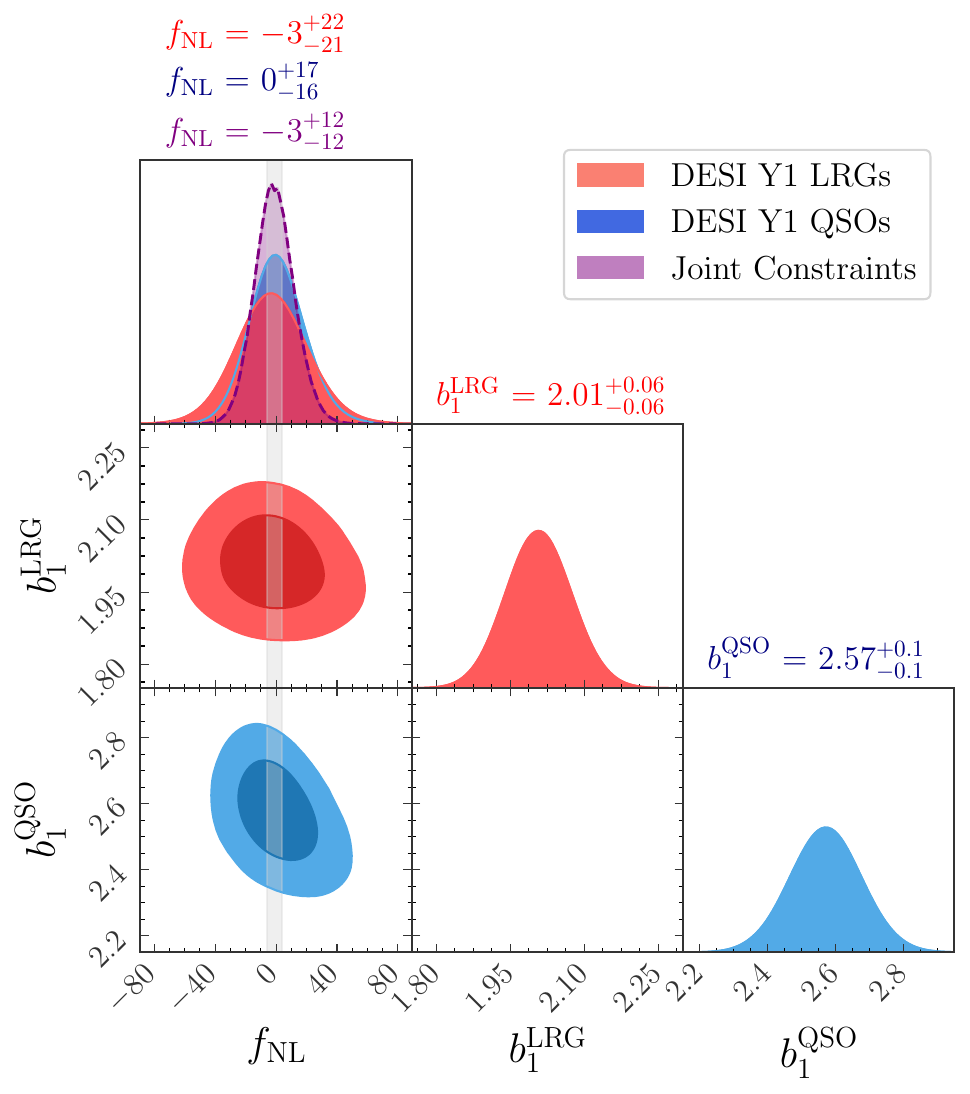}
\caption{Marginalized distributions of the POIs for the LRG sample (red), QSO sample (blue), and an $f_{\mathrm{NL}}$ posterior for the joint fit (purple). The grey shaded band denotes the 1$\sigma$ uncertainties from the most precise Planck CMB measurement to date. The dark and light regions represent $1\sigma$ and $2\sigma$ contours respectively.}
\label{fig:post_comb}
\end{figure}

\begin{figure}
\includegraphics[width=\columnwidth]{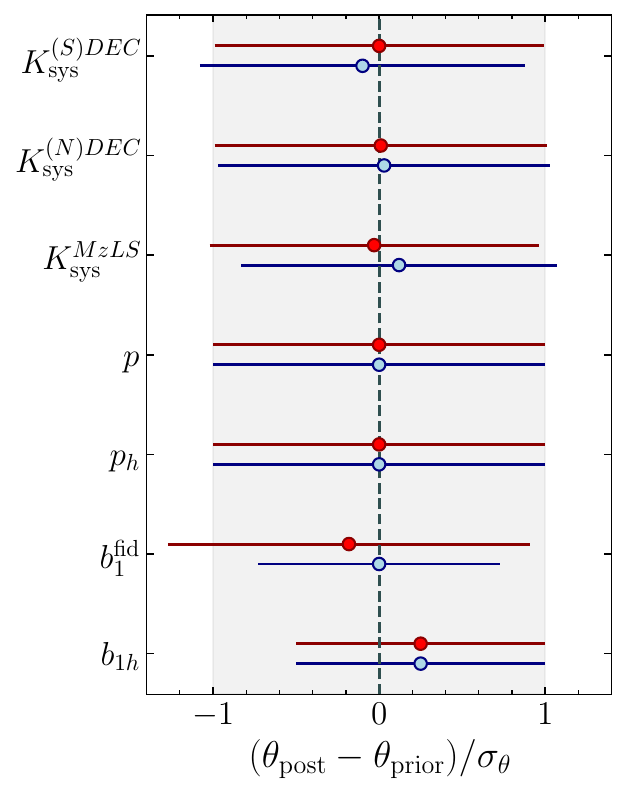}
\caption{The size of our constraints on the nuisance parameters relative to their priors, $( \theta_{\rm post} - \theta_{\rm prior} )/\sigma_{\theta}$ for both the LRG (red) and QSO (blue) chains. The grey shaded region indicates the 1$\sigma$ prior width.}
\label{fig:sys_budget}
\end{figure}

To leverage the constraining power of both samples simultaneously we combine the LRG and QSO posteriors, and perform a joint fit using both tracer samples. We present this measurement in Fig.~\ref{fig:post_comb}. The combination of DESI DR1 LRGs and QSOs yields 1$\sigma$ uncertainties on $f_{\mathrm{NL}}$ of approximately 12. 
We can conclude that within the chosen fiducial cosmology, the DESI DR1 samples prefers a value of $f_{\mathrm{NL}}$ which is consistent with zero.

We investigate the systematics budget of our pipeline by comparing the size of the posterior to the prior distributions of the nuisance parameters. We show this comparison in Fig.~\ref{fig:sys_budget}. In the case of the imaging systematics and $p$ parameters, they are entirely set by the priors (see Tab.~\ref{tab:priors}). The size of the resultant $f_{\rm NL}$ constraints are naturally coupled to the choice of $p$-prior (both width and central value). We have found that imposing a strict prior on $p$ (the value is known exactly) could potentially reduce the size of the $f_{\rm NL}$ constraints by $\sim$$5 \%$ for both LRGs and QSOs. However, we feel strongly that allowing a small variation in the fiducial $p$-priors more accurately reflects our knowledge of the model parameters.

For the linear bias of the FastPM halo simulations, both LRGs and QSOs are able to constrain the value of $b_{1h}$ to better precision than the prior. For the fidicual linear bias, $b_1^{\rm fid}$, only the QSO case shows constraining power beyond the chosen prior.


\section{Discussion}
\label{sec:discussion}

This work has established, in our estimation, three important conclusions regarding PNG studies in LSS. The first, which was also highlighted in \cite{brown2024constrainingprimordialnongaussianitylarge}, is that a robust PNG study should be carried out in both Fourier-space and configuration-space. The effect of PNG on the power spectrum is highly localized to small $k$-modes, while in configuration-space, it is spread out to a broader range of scales. Thus, the interplay between systematics and PNG is very different. We do not claim that configuration-space clustering is a better probe of PNG, simply that a robust approach must involve both. For DESI DR1, the Fourier space $f_{\mathrm{NL}}$ constraints presented in \cite{chaussidon2025constrainingprimordialnongaussianitydesi} are consistent with the ones presented here. 

Secondly, we wish to emphasize the importance of the mitigation of imaging systematics for any LSS PNG measurement. Especially in cases such as DESI DR1, where large angular density variations exist due to survey incompleteness, this is critical. We find the aforementioned consistency of this study with Fourier-space fits particularly encouraging, since the two methods employ independent methods of systematics mitigation. 

Finally, and most importantly, this work definitively demonstrates the constraining power of LSS surveys. With only the first DESI data release, we are able to achieve constraints approaching the sensitivity of CMB measurements.

A potentially unmitigated systematic is associated with the periodic nature of the fiducial model based on AbacusSummit, where the original simulation box of (2 $h^{-1}$ Gpc)$^3$ must be tiled to produce the irregular DESI cutsky catalog. Since the maximum scale probed in this analysis is 400 $h^{-1}$ Mpc we expect a negligible effect on the final result. However, in the future versions of this analysis, it is prudent to use simulations of larger volumes.


\section{Conclusions}
\label{sec:conclusions}

In this work we have presented measurements of $f_{\mathrm{NL}}$ using the 2pcf of DESI DR1 LRGs and QSOs. For both tracers, and the combined fit, we find that the data prefer a value of $f_{\mathrm{NL}}=-3 \pm 12$, which is consistent with zero. This result must interpreted in the context of the corresponding Gaussian $p$-priors, $p_{\rm LRG} = 1.0\pm 0.1$ and $p_{\rm QSO} = 1.6\pm 0.1$. Extending the lessons we've learned with DESI DR1 tracers, we believe that we have now entered an era where the most sensitive probes of primordial physics will be the characterization of cosmological structure at late times.


\section*{Acknowledgments}

This material is based upon work supported by the U.S. Department of Energy (DOE), Office of Science, Office of High-Energy Physics, under Contracts No. DE–AC02–05CH11231,  DE-SC0008475, and by the National Energy Research Scientific Computing Center, a DOE Office of Science User Facility under the same contract. 

Additional support for DESI was provided by the U.S. National Science Foundation (NSF), Division of Astronomical Sciences under Contract No. AST-0950945 to the NSF’s National Optical-Infrared Astronomy Research Laboratory; the Science and Technology Facilities Council of the United Kingdom; the Gordon and Betty Moore Foundation; the Heising-Simons Foundation; the French Alternative Energies and Atomic Energy Commission (CEA); the National Council of Science and Technology of Mexico (CONACYT); the Ministry of Science and Innovation of Spain (MICINN), and by the DESI Member Institutions: \url{https://www.desi.lbl.gov/collaborating-institutions}. 

Any opinions, findings, and conclusions or recommendations expressed in this material are those of the author(s) and do not necessarily reflect the views of the U. S. National Science Foundation, the U. S. Department of Energy, or any of the listed funding agencies. 

The authors are honored to be permitted to conduct scientific research on Iolkam Du’ag (Kitt Peak), a mountain with particular significance to the Tohono O’odham Nation.

\section*{Data Availability}

All galaxy and quasar samples used in this study are available as part of the Dark Energy Spectroscopic Survey's Year-One data release. Please contact the authors for access to the collection of Fast-PM simulations. The data shown in this paper's figures are available in machine-readable format at \url{https://zenodo.org/records/20751024}.


\bibliographystyle{mnras}
\bibliography{bib} 



\appendix

\section{Author Affiliations}
\label{app:affiliations}

{\it
$^{1}$ Department of Physics, Kansas State University, 116 Cardwell Hall, Manhattan, KS 66506, USA \\
$^{2}$Department of Physics and Astronomy, University of Rochester, 500 Joseph C. Wilson Boulevard, Rochester, NY 14627, USA \\
$^{3}$ Lawrence Berkeley National Laboratory, 1 Cyclotron Road, Berkeley, CA 94720, USA \\
$^{4}$ Department of Physics, Boston University, 590 Commonwealth Avenue, Boston, MA 02215 USA \\
$^{5}$ Dipartimento di Fisica ``Aldo Pontremoli'', Universit\`a degli Studi di Milano, Via Celoria 16, I-20133 Milano, Italy \\
$^{6}$ INAF-Osservatorio Astronomico di Brera, Via Brera 28, 20122 Milano, Italy \\
$^{7}$ Department of Physics \& Astronomy, University College London, Gower Street, London, WC1E 6BT, UK \\
$^{8}$ Instituto de F\'{\i}sica, Universidad Nacional Aut\'{o}noma de M\'{e}xico, Circuito de la Investigaci\'{o}n Cient\'{\i}fica, Ciudad Universitaria, Cd. de M\'{e}xico C.~P.~04510, M\'{e}xico \\
$^{9}$ Department of Astronomy \& Astrophysics, University of Toronto, Toronto, ON M5S 3H4, Canada \\
$^{10}$ Department of Physics \& Astronomy and PITT PACC, University of Pittsburgh, 3941 O'Hara Street, Pittsburgh, PA 15260, USA \\
$^{11}$ Departamento de F\'isica, Universidad de los Andes, Cra. 1 No. 18A-10, Edificio Ip, CP 111711, Bogot\'a, Colombia \\
$^{12}$ Observatorio Astron\'omico, Universidad de los Andes, Cra. 1 No. 18A-10, Edificio H, CP 111711 Bogot\'a, Colombia \\
$^{13}$ Institut d'Estudis Espacials de Catalunya (IEEC), c/ Esteve Terradas 1, Edifici RDIT, Campus PMT-UPC, 08860 Castelldefels, Spain \\
$^{14}$ Institute of Cosmology and Gravitation, University of Portsmouth, Dennis Sciama Building, Portsmouth, PO1 3FX, UK \\
$^{15}$ Institute of Space Sciences, ICE-CSIC, Campus UAB, Carrer de Can Magrans s/n, 08913 Bellaterra, Barcelona, Spain \\
$^{16}$ University of Virginia, Department of Astronomy, Charlottesville, VA 22904, USA \\
$^{17}$ Fermi National Accelerator Laboratory, PO Box 500, Batavia, IL 60510, USA \\
$^{18}$ Department of Astronomy, University of Texas at Austin, 2515 Speedway, TX 78712, USA \\
$^{19}$ Center for Cosmology and AstroParticle Physics, The Ohio State University, Columbus, OH 43210, USA \\
$^{20}$ Department of Physics, The Ohio State University, Columbus, OH 43210, USA \\
$^{21}$ The Ohio State University, Columbus, OH 43210, USA \\
$^{22}$ Department of Physics, University of Michigan, 450 Church Street, Ann Arbor, MI 48109, USA \\
$^{23}$ University of Michigan, 500 S. State Street, Ann Arbor, MI 48109, USA \\
$^{24}$ Department of Physics, The University of Texas at Dallas, 800 W. Campbell Rd., Richardson, TX 75080, USA \\
$^{25}$ NSF NOIRLab, 950 N. Cherry Ave., Tucson, AZ 85719, USA \\
$^{26}$ Department of Physics and Astronomy, University of California, Irvine, CA 92697, USA \\
$^{27}$ Departament de F\'{i}sica, Universitat Aut\`{o}noma de Barcelona, 08193 Bellaterra (Barcelona), Spain \\
$^{28}$ Institut de F\'{i}sica d'Altes Energies (IFAE), The Barcelona Institute of Science and Technology, Campus UAB, 08193 Bellaterra (Barcelona), Spain \\
$^{29}$ Instituci\'{o} Catalana de Recerca i Estudis Avan\c{c}ats, Passeig de Llu\'{\i}s Companys 23, 08010 Barcelona, Spain \\
$^{30}$ Department of Physics and Astronomy, University of Waterloo, Waterloo, ON N2L 3G1, Canada \\
$^{31}$ Perimeter Institute for Theoretical Physics, Waterloo, ON N2L 2Y5, Canada \\
$^{32}$ Waterloo Centre for Astrophysics, University of Waterloo, Waterloo, ON N2L 3G1, Canada \\
$^{33}$ Departament de F\'isica, EEBE, Universitat Polit\`ecnica de Catalunya, c/Eduard Maristany 10, 08930 Barcelona, Spain \\
$^{34}$ Department of Physics and Astronomy, Sejong University, Seoul 05006, Republic of Korea \\
$^{35}$ Abastumani Astrophysical Observatory, Tbilisi, GE-0179, Georgia \\
$^{36}$ Ilia State University, Tbilisi, 0194, Georgia \\
$^{37}$ CIEMAT, Avenida Complutense 40, E-28040 Madrid, Spain \\
$^{38}$ National Astronomical Observatories, Chinese Academy of Sciences, Beijing 100101, P.~R.~China \\
$^{39}$ Instituto de F\'{\i}sica Te\'{o}rica (IFT) UAM/CSIC, Universidad Aut\'{o}noma de Madrid, Cantoblanco, E-28049, Madrid, Spain \\
$^{40}$ Institut de F\'{i}sica d'Altes Energies (IFAE), The Barcelona Institute of Science and Technology, Campus UAB, 08193 Bellaterra Barcelona, Spain \\
$^{41}$ Centro de Investigaci\'{o}n Avanzada en F\'{\i}sica Fundamental (CIAFF), Facultad de Ciencias, Universidad Aut\'{o}noma de Madrid, ES-28049 Madrid, Spain \\
}

\section{Masking BAO Scales}
\label{app:bao_mask}

\begin{figure}
\includegraphics[width=\columnwidth]{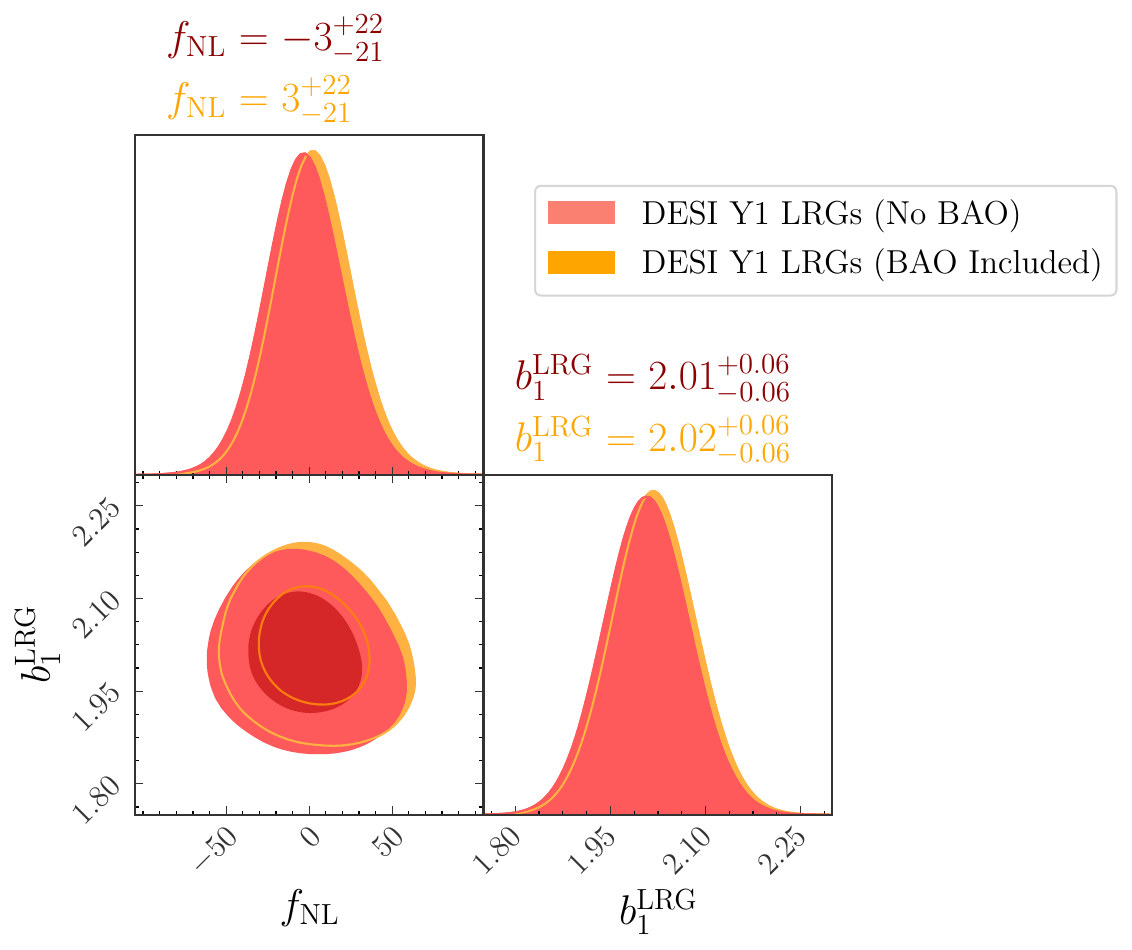}
\caption{Marginalized distributions of the POIs for DESI Y1 LRGs over the full range of scales (orange) including the BAO, and by removing the BAO scales (red). The dark and light regions represent $1\sigma$ and $2\sigma$ contours, respectively. The most probable values and $1\sigma$ CL are labeled for each parameter above the respective panel. }
\label{fig:win_lrg}
\end{figure}

\begin{figure}
\includegraphics[width=\columnwidth]{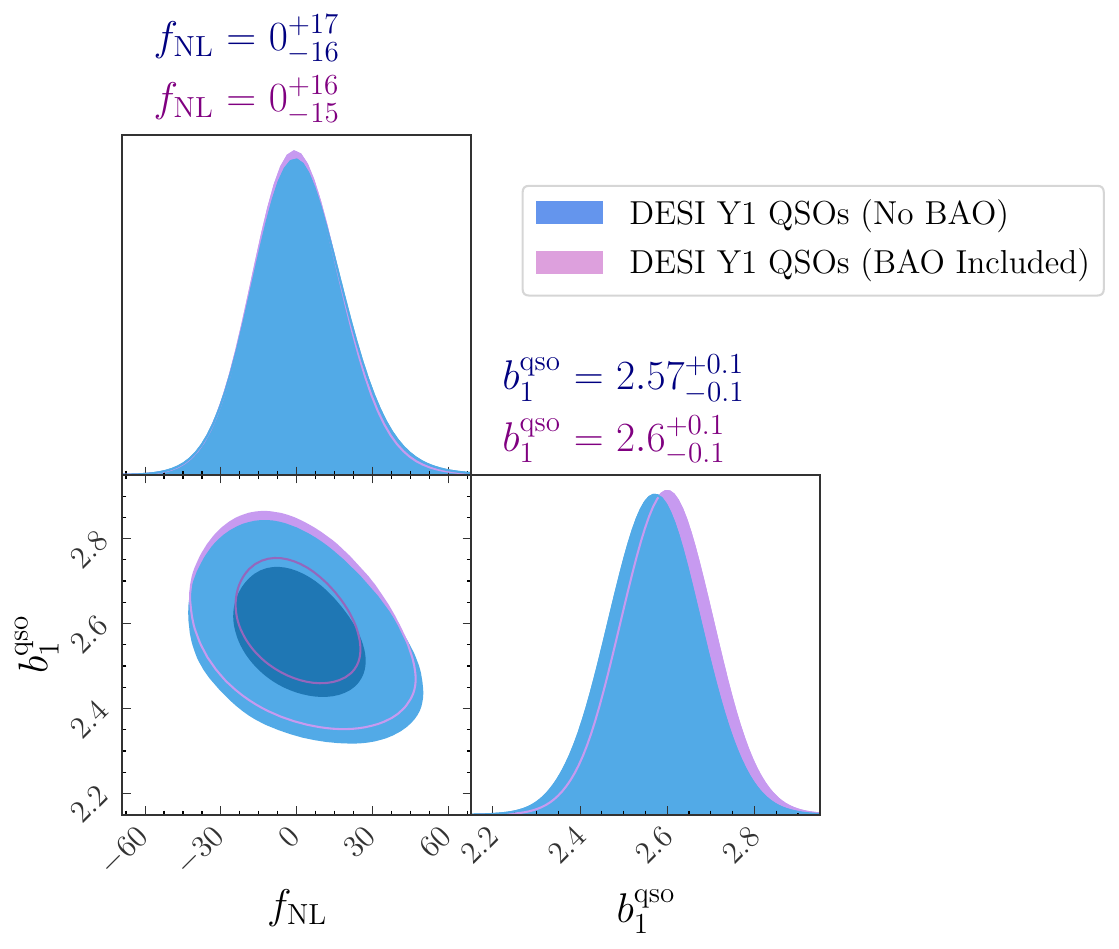}
\caption{The same as Fig.~\ref{fig:win_lrg} for DR1 QSOs over the full range of scales (purple) and by removing the BAO scales (blue).}
\label{fig:win_qso}
\end{figure}

In our model, we employ Gaussian priors on the value of the linear bias corresponding to the used simulations, $b_1^{\mathrm{fid}}$, and the PNG halo simulations, $b_{1h}$. To inform these priors, we directly measure the bias of the simulated tracers using a simple approach described in \cite{brown2024constrainingprimordialnongaussianitylarge}. It creates a template 2pcf by transforming the linear power spectrum. The clustering properties of our simulations, however, are non-linear. For the scales considered in our analysis (50 $h^{-1}$Mpc $<$ $s$ $<$ 380 $h^{-1}$Mpc), this effect is largest near the BAO peak.

We test the effect that BAO scale clustering has on our measurement by removing the mask (100 $h^{-1}$Mpc $<$ $s$ $<$ 120 $h^{-1}$Mpc) and repeating the procedure. The resultant $f_{\mathrm{NL}}$ constraints are shown in Fig.~\ref{fig:win_lrg} for the LRGs, and in Fig.~\ref{fig:win_qso} for the QSOs. In both cases we find little to no difference in the $f_{\mathrm{NL}}$ contours. From these tests, we remain satisfied by the robustness of our pipeline even in the presence of some non-linear clustering near the BAO scale. However, since we know the BAO scales capture clustering not included in our modelling, we still exclude them from the fits presented in the main body of this paper out of an abundance of caution.


\bsp	
\label{lastpage}
\end{document}